\documentclass[11pt,preprint]{aastex}
\usepackage{apjfonts}
\usepackage{placeins}

\shorttitle{Rapid accretion onto protostars} \shortauthors{Hartmann et al.}

\begin{document}

\title{On Rapid Disk Accretion and Initial Conditions in Protostellar Evolution}

\author{Lee Hartmann\altaffilmark{1}, Zhaohuan Zhu\altaffilmark{1}, 
\& Nuria Calvet\altaffilmark{1}}

\altaffiltext{1}{Dept. of Astronomy, University of Michigan, 500
Church St., Ann Arbor, MI 48105} 
\email{lhartm@umich.edu}

\newcommand\msun{M_{\odot}}
\newcommand\lsun{L_{\odot}}
\newcommand\msunyr{M_{\odot}\,yr^{-1}}
\newcommand\be{\begin{equation}}
\newcommand\en{\end{equation}}
\newcommand\cm{\rm cm}
\newcommand\kms{\rm{\, km \, s^{-1}}}
\newcommand\K{\rm K}
\newcommand\etal{{et al}.\ }
\newcommand\sd{\partial}
\newcommand\mdot{\dot{M}}
\newcommand\rsun{R_{\odot}}
\newcommand\yr{\rm yr}

\begin{abstract}
Low-mass protostars may accrete most of their material through short-lived
episodes of rapid disk accretion; yet until recently evolutionary tracks for
these protostars assumed only constant or slowly-varying accretion.
Important initial steps toward examining the potential effects of rapid 
accretion were recently made by Baraffe, Chabrier, \& Gallardo,
who showed that in the limit of low-temperature ("cold") 
accretion, protostars may have much smaller radii than found 
in previous treatments.  Such small radii at the end of protostellar
accretion would have the effect of
making some young stars appear much older - perhaps as much as 10 Myr -
than they really are.
However, we argue that very rapid disk accretion is unlikely to be cold,
for two reasons.  First, observations of the 
best-studied pre-main sequence disks with rapid disk accretion outbursts 
- the FU Ori objects -  have spectral energy distributions
which imply large, not small, protostellar radii.  Second, theoretical
models indicate that at high accretion rates, protostellar disks become 
internally hot and geometrically thick, making it much more likely that 
hot material is added to the star.  In addition, the very large luminosity 
of the accretion disk is likely to irradiate the central star strongly, 
heating up the outer layers and potentially expanding them.  Finally,
observed HR diagram positions of most young stars and even estimated properties
of some Class 0 protostars are inconsistent with the rapid cold accretion models.
Nevertheless, the Baraffe et al. calculations emphasize the importance of
initial protostellar radii for subsequent evolution.
As these initial properties are controlled by the amount of thermal energy incorporated
into the protostellar core during initial hydrodynamic collapse,
significant variations in initial protostellar radii for a given final
stellar mass are quite possible if not likely.
These variations could well affect the apparent ages of low-mass stars near the nominal
1 Myr isochrones.
\end{abstract}


\section{Introduction}
Isochrones for pre-main sequence stars are of great importance for our
understanding of early stellar evolution, histories of star formation in
molecular clouds, and evolutionary timescales of protoplanetary disks.
The contraction of these non-hydrogen-burning stars can be used to estimate
their ages, provided the ``initial'' radius at the end of protostellar accretion is
known.  The starting radius is important for age-dating the youngest
stars; theoretical studies long ago indicated that solar-mass protostars
have radii only a few times larger than their main sequence values (Larson 1969, 1972; 
Stahler, Shu, \& Taam 1980a,b; Stahler 1983, 1988), a result which has been 
reinforced over the years (Palla \& Stahler 1992, 1999; Omukai \& Palla 2003; 
Hosokawa \& Omukai 2009). 

The protostellar radius is sensitive to the exact amount of thermal energy 
added during accretion (in the absence of nuclear fusion), as gas pressure 
provides the support against gravity.  Fully hydrodynamic collapse calculations 
from realistic turbulent and time-dependent initial conditions
usually assume simplistic energy equations (e.g., Klessen, Burkert, \& Bate 1998; 
Bate, Bonnell, \& Bromm 2003), because of the large range of scales 
that must be followed (for example, the recent calculations of Greif et al.
2011 follow collapse down to accretion radii of 100 $\rsun$).
Therefore most calculations of the detailed evolution of protostellar
properties have assumed some initial protostellar core mass 
and radius with some assumed infall rate to resolve the accretion 
flow - and thus, the heat transport - much more finely.  

The most self-consistent treatments of thermal energy addition during accretion
have assumed spherical (or quasi-spherical) infall 
(Stahler \etal 1980a,b; Stahler 1983; Hosokawa \& Omukai 2009).
However, it seems likely that much, if not most of the protostellar mass
must be accreted through a disk, rather than spherically, given the
small sizes of protostars relative to their progenitor cloud and
the likelihood that the cloud had some finite angular momentum.  This greatly
increases the difficulty in calculating the amount of thermal energy 
transported into the protostar.  For one thing, the problem can no longer be treated
in spherical geometry; for another, the details of the disk structure depend upon
the nature and properties of the viscous transport, which are uncertain
(see, e.g., Hartmann \etal 2006).  To get around this difficulty,
Siess, Forestini, \& Bertout (1997, 1999; SFB) and Hartmann, Cassen, \& Kenyon (1997; HCK) 
treated the amount of thermal energy added to the star via disk accretion
as a free parameter.  SFB and HCK further showed that for modest accretion
rates, the assumption of negligible thermal energy addition (``cold accretion'') was  
arguably adequate.  Moerover, these results were 
roughly consistent with the spherical accretion calculations of Stahler (1988),
basically because the onset of D fusion lead to protostellar evolutionary
tracks near what might be termed the ``D-fusion main sequence'' 
(important in establishing the protostellar ``birthline'' (end of major accretion)
positions in the HR diagram; Stahler 1983, 1988).
The above investigations led to birthlines in reasonable agreement with observations,
in that they were situated modestly above the observed positions of pre-main sequence stars
in the HR diagram, consistent with the expectation that the stars contract 
toward the main sequence after the main phase of accretion has ended.  
In addition, recent improvements in directly estimating protostellar radii
during the Class I phase also show reasonable agreement with theory 
(White \etal 2007, and references therein).

These investigations of protostellar properties generally assumed 
steady or slowly-varying accretion as a function of time.  However,
there is increasing evidence that low-mass protostars accrete a substantial
part, if not most, of their material through short bursts of extremely
rapid disk accretion (Kenyon \etal 1990, 1994; Enoch \etal 2009; Dunham \etal 2010),
with extended periods of low mass accretion interspersed (Muzerolle \etal 1998;
White \& Hillenbrand 2004; Doppmann \etal 2005; Covey \etal 2005).  
Motivated by these considerations, Baraffe, Chabrier, \& Gallardo 
(2009; BCG) and Baraffe \& Chabrier (2010; BC) made an important first step by
calculating models with bursts of rapid protostellar accretion.  Because of the
major difficulties involved in computing the thermal energy added through disk
accretion described above, BCG and BC adopted the limiting case of cold accretion.
BCG and BC showed that it was possible to obtain much smaller birthline radii 
than found in previous studies, even lying well below many of the HR diagram
positions of young populations in molecular clouds. 

Given the potential significance of the BCG and BC results for determining
ages of young stars, we attempt here to address the question: 
can such rapid disk accretion be treated in the cold limit?
We suggest that the answer is no, based on both observational
and theoretical considerations.  However, the BCG and BC results
{\em also} depend on the assumption of a small initial radius.
As some variation in initial protostellar radii must occur, depending upon
the specific circumstances in which individual stars form,
the BCG and BC investigations serve as a reminder of the potential importance
of initial conditions (a point made previously by HCK).
Based on current observations and limitations on the possibility of rapid
cold accretion, we argue that while uncertainties of order 10 Myr are
very unlikely, uncertainties of $\sim$~1 Myr due to variations in
initial protostellar core radii seem probable.

\section{An observational test}

\begin{figure}
\includegraphics[angle=0,width=1.1\textwidth]{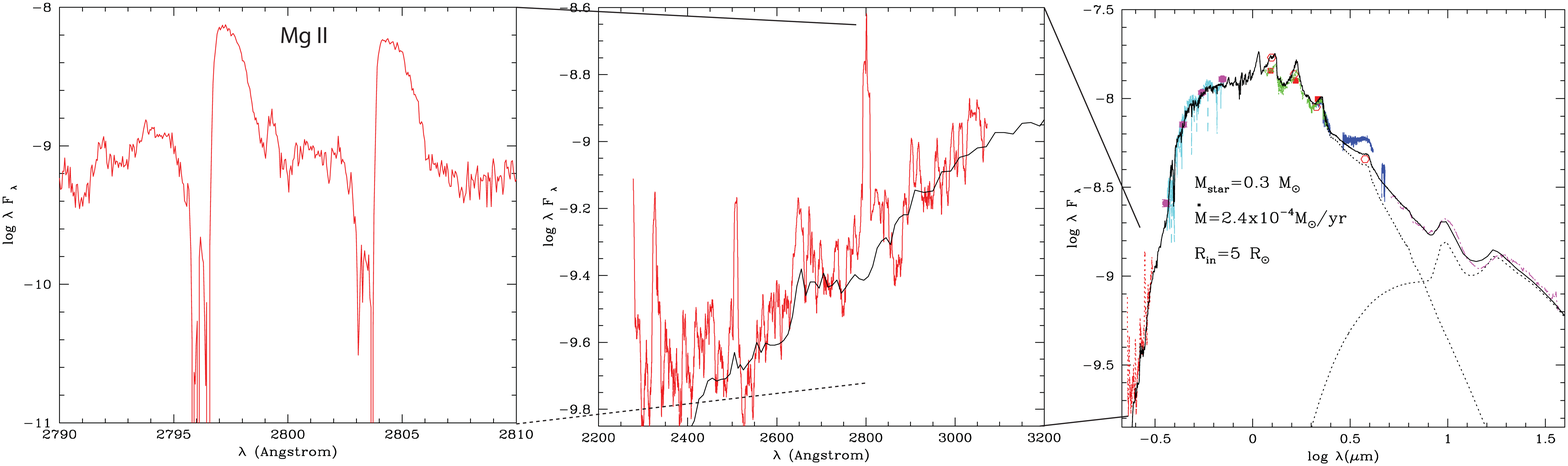}
\caption{Observed spectral energy distribution of FU Ori,
compared with the theoretical disk model of Zhu \etal (2008)
(middle and right panels), with details of the Mg II resonance
line profile (left panel).
The data are taken from the literature as indicated in Zhu \etal
(2008) with the addition of an ultraviolet STIS spectrum from
the HST archive (ads/Sa.HST\#O63L07010), Calvet PI. 
The red curves show the observations and the black solid curve
the model prediction.
The model has not been adjusted to fit the ultraviolet spectrum.
The overall fit indicates the appropriateness of
the adopted maximum steady disk effective temperature and the
absence of any measurable boundary layer radiation (see text).}
\label{fig:fig1}
\end{figure}

The outbursts of FU Ori objects result from disk accretion at
$\mdot \sim 10^{-4} \msunyr$ (Hartmann \& Kenyon 1996), similar to
the rates assumed in BCG and BC.  In Figure \ref{fig:fig1} we show a revised version
our recent fit for spectral energy distribution (SED) of FU Ori itself 
using a steady disk model (Zhu \etal 2007, 2008; Figure \ref{fig:fig1}). 
Here we use the HST STIS spectrum (Project 8627, Calvet PI) to extend results 
to the near-ultraviolet {\em without changing the model or estimated extinction}.
As can be seen, the simple steady disk model fits the SED very well from
$\sim 2500$~\AA\ to $\sim 8 \mu$m, beyond which the spectrum changes from that
of an internally-heated viscous disk to that of a flared disk heated mostly
by radiation from the inner regions.

Using our measurements of Keplerian rotation from $0.6 \mu$m to $4.6\mu$m 
(Zhu \etal 2009a) and the estimated inclination
of $i = 55^{\circ}$ from near-infrared interferometry, the
central star mass is $\simeq 0.3 \msun$, the inner disk radius is $\simeq 5 \rsun$,
and the accretion rate is $\mdot \simeq 2 \times 10^{-4} \msunyr$.
This radius, though consistent with previous estimates for other
well-studied FU Ori objects (e.g., Kenyon, Hartmann, \& Hewett 1988; KHH), 
differs strongly from the low-luminosity models of BC and BCG.  For instance,
BC find a radius of $\sim 0.9 \rsun$ in their Figure 5 for a $0.3 \msun$ protostar
which has undergone rapid cold accretion.  As the maximum disk temperature 
scales as $L_{acc}^{1/4} R_*^{-1/2}$, for the same total luminosity
an inner disk radius of $1 \rsun$ would change the
peak disk temperature from our model result of $T_m \sim 6500$~K to 
$\sim 14,000$~K.  The optical-near UV spectral features, which are independent
of accretion, would then have the appearance of a late B star rather than the
observed early G star, well beyond any observational uncertainties.
Indeed, the inferred inner radius is also larger than that of typical birthline
estimates, leading to suggestions that rapid disk accretion leads to
stellar {\em expansion}, not contraction (Hartmann \& Kenyon 1985; Kenyon \etal 1989).

One way of potentially reconciling this observational result with a smaller central
star radius is if a stellar magnetosphere is truncating the disk.  However, as shown
in the middle panel of Figure \ref{fig:fig1}, there is no evidence for accretion
shock emission at a level $\gtrsim 10^{-2}$ of the overall accretion luminosity.
Furthermore, the observed line profiles show no evidence for magnetospheric infall,
as discussed in Hartmann \& Kenyon (1996); the major emission lines simply show
outflow, as in the case of the Mg II resonance lines, with blue-shifted absorption
(left panel of Figure \ref{fig:fig1}). 
Indeed, the absence of a noticeable high-temperature component
either as a result of magnetospheric accretion shock or boundary layer emission 
poses some interesting theoretical questions, as discussed in the following section.

\section{Rapidly-accreting disks}

The accretion luminosity (ignoring boundary layer emission) of a (steady) disk is
\begin{equation}
L_{acc} ~=~ {G M_* \dot{M} \over 2 R_*} \sim 1.5 \times 10^3 \left ({M_* \over \msun} \right)
\left ( {\mdot \over 10^{-4} \msunyr} \right) 
\left( {\rsun \over R_*} \right) \, \lsun \,.
\end{equation}
Initially, the masses of of the BCG and BC objects are quite low, $\sim 10^{-3}
\msun$, with radii on the order of $10^{-1} \rsun$; at this stage, the intrinsic (stellar)
luminosities
are also low, $10^{-2} \lsun$ or smaller.  Later on, for protostellar masses of $\sim 0.1 -0.3
\msun$ and radii $\sim 1 \rsun$, luminosities are $\sim 0.1 \lsun$.  
Thus, during the presumed cold rapid accretion
phase, the accretion luminosities are three orders of magnitude larger than
that of the (sub)stellar central object.  For cold accretion to be a reasonable
approximation, it is therefore necessary for only $10^{-3}$ or less of the
accretion energy to enter the star as thermal energy; at larger rates,
expansion of the star is likely (see, e.g., Prialnik \& Livio 1985).
This is a very stringent requirement, for the following reasons.

\begin{figure}
\includegraphics[angle=0,width=0.7\textwidth]{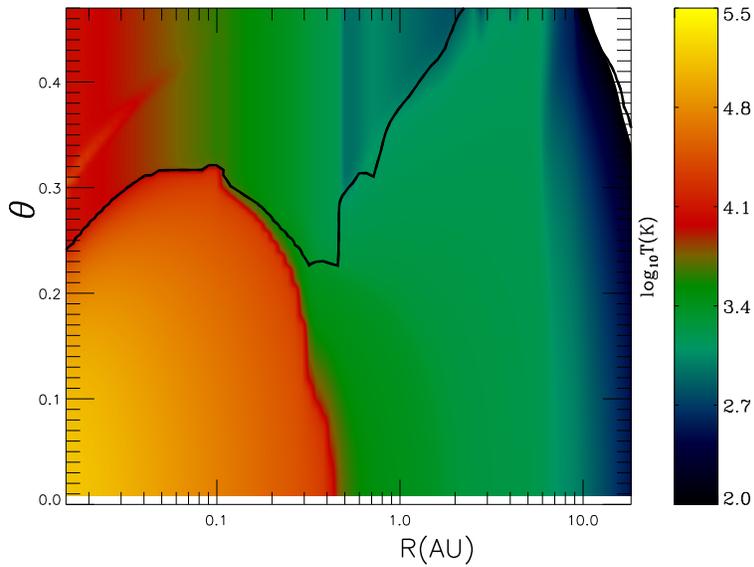}
\caption{Cross-section of the inner region of
a disk accreting at $10^{-4} \msunyr$
with a 1 $\msun$ central star.  The angle
$\theta$ is the polar angle in radians measured from
the disk midplane and $R$ is the radial distance
from the star (see Zhu \etal 2009c for details).  The midplane temperature
exceeds $1 \times 10^5$~K at small radii.  The estmated disk
photosphere, indicated by the black curve,
extends to $\sim 0.3$~rad from the midplane.
(The outer disk photosphere is uncertain because of
limitations of the boundary conditions used in the two-dimensional
simulations.)}
\label{fig:fig2}
\end{figure}

Models have long indicated that FU Ori disks should have very 
high midplane temperatures at small radii, $T_c \gtrsim 1 \times 10^5$~K, 
and thus should become geometrically thick 
(Clarke, Lin, \& Pringle 1990; Bell \& Lin 1994;
Zhu \etal 2009c; Figure \ref{fig:fig2}).  As Clarke \etal (1990) pointed out, this geometrical 
thickness helps explain the absence of boundary layer radiation; where in
standard disk models slowing from Keplerian rotation occurs over a narrow 
radial region, resulting in high effective temperatures, the diffusion of this energy
over a radial region $\Delta R \sim H$ eliminates this emission
(see also Popham \etal 1996).  Indeed, the 
steady disk model without boundary layer emission fits the HST spectrum remarkably
well (Figure \ref{fig:fig1}).  The absence of hot emission, either from a boundary layer
or a magnetospheric accretion shock, led Hartmann \& Kenyon (1985) and Kenyon \etal 
(1989) to suggest that significant amounts of thermal energy from the disk are
being more widely distributed, not just within the disk but into the star.

It is worth emphasizing that the advection of hot material into the star is not
the only way in which rapid accretion can heat the central protostar.  Simply by
virtue of being geometrically thick, the disk blocks a significant fraction of
the protostellar surface, so that a freely-radiating boundary condition is no
longer appropriate.  Perhaps more importantly, some of the radiation emitted
by the disk can be absorbed by the stellar outer layers.  While for low
accretion rates and geometrically thin disks such heating is relatively
unimportant (e.g., Adams \& Shu 1986), this is probably not the case when
$L_{acc} \gtrsim 10^3 L_*$. 

The radius interior to which the disk is
expected to make the transition to a state with a high central temperature, and thus
become geometrically thick, is (see Appendix):
\begin{equation}
R_T \sim 13 \left ( {M \over 0.3 \msun}\right)^{0.33}
\left ( {\mdot \over 10^{-5} \msunyr} \right)^{0.42}\,
\left ( {\alpha \over 0.1}\right)^{-0.25}\, \rsun \,,
\label{eq:R(mdot)}
\end{equation}
where $M$ is the central star mass, $\mdot$ is the (assumed steady disk)
accretion rate in units of
$10^{-5} \msunyr$, and $\alpha$ is the usual viscosity parameter.
The prediction of equation (\ref{eq:R(mdot)}) compares well with the detailed numerical
simulations of Zhu \etal (2009b) at higher accretion rates.
This result reinforces the conclusion that
the assumption of thin cold disk accretion is unlikely
to be reasonable at $\mdot \sim 10^{-4} \msunyr$, 
and there may be effects even at accretion rates of $10^{-5} \msunyr$ onto a 
low-mass protostars. 

It is plausible that even if the stellar magnetosphere is crushed down to the
stellar surface at FU Ori accretion rates, at lower mass fluxes the magnetosphere could
hold off the disk from the stellar photosphere.  However, even in this case,
if the disk outside the magnetosphere is geometrically thick as indicated by
equation (\ref{eq:R(mdot)}), some accretion energy will be radiated out of the
inner disk ``wall'' at the magnetospheric truncation radius; for
$L_{acc} \gg L_*$, very little of this radiation need be intercepted
by the star to change the outer boundary condition radically.  We conjecture
that this would inevitably heat the outer layers of the star significantly even
in the absence of direct advection of thermal energy by mass accretion.

\section{Initial radii}

While BCG and BC assumed cold accretion occurred during disk accretion outbursts 
(Vorobyov \& Basu 2005, Vorobyov 2009, Zhu \etal 2009 a,b,c), BCG note 
that outbursts are not necessary to produce small protostellar radii; rather, the
radius at the onset of the calculation was crucial.
To see qualitatively why the initial radius is so important,
we construct the following highly simplified model.  Following HCK, we model
the protostar as an $n = 3/2$ polytrope, assume cold accretion, and neglect any
blocking of the stellar photospheric radiation by optically-thick accreting
material.  The evolution of the protostar is then determined by the equation
(HCK)
\begin{equation}
L_* ~=~ - {3 \over 7} {G M_*^2 \over R_*}
\left [ \left ( {1 \over 3} ~-~ {7 \alpha_A \over 3} \right ) {\mdot \over M_*}
~+~ { \dot{R_*} \over R_*} \right ] ~+~ L_D \,. \label{eq:laccd}
\end{equation}
Here $\alpha_A$ parameterizes the thermal energy addition per unit mass, and
the deuterium fusion luminosity is given by
\begin{equation}
L_D ~=~ 1.92 \times 10^{17}
\,  f  \, [D/H] \,
\left ( {M_* \over \msun} \right )^{13.8} \,
\left ( {R_* \over \msun} \right )^{-14.8}\, \lsun \label{eq:ld}
\end{equation}
(Stahler 1988). Here $f$ is the fractional concentration of deuterium relative to
its initial number abundance, which is taken to be $[D/H] = 2.5 \times 10^{-5}$.

We evolve the protostellar properties using equation (\ref{eq:laccd}) for
$\alpha_A = 0$ for high mass accretion rates that are constant in time.  The
method of solution is essentially the same as in HCK, except that we adopt a
simple estimate of the effective temperature as an approximation to the
Baraffe \etal (1998) tracks near $1-3$~Myr between $\sim 0.1 - 1.4 \msun$:
\begin{equation}
\log T_{eff} ~=~ 3.48 ~+~ 0.147 (M/\msun)\,.
\end{equation}
This essentially assumes that the Hayashi tracks are vertical in the 
HR diagram.  While this is not quite correct, the results are not very sensitive
to this simplification.

\begin{figure}
\includegraphics[angle=0,width=0.7\textwidth]{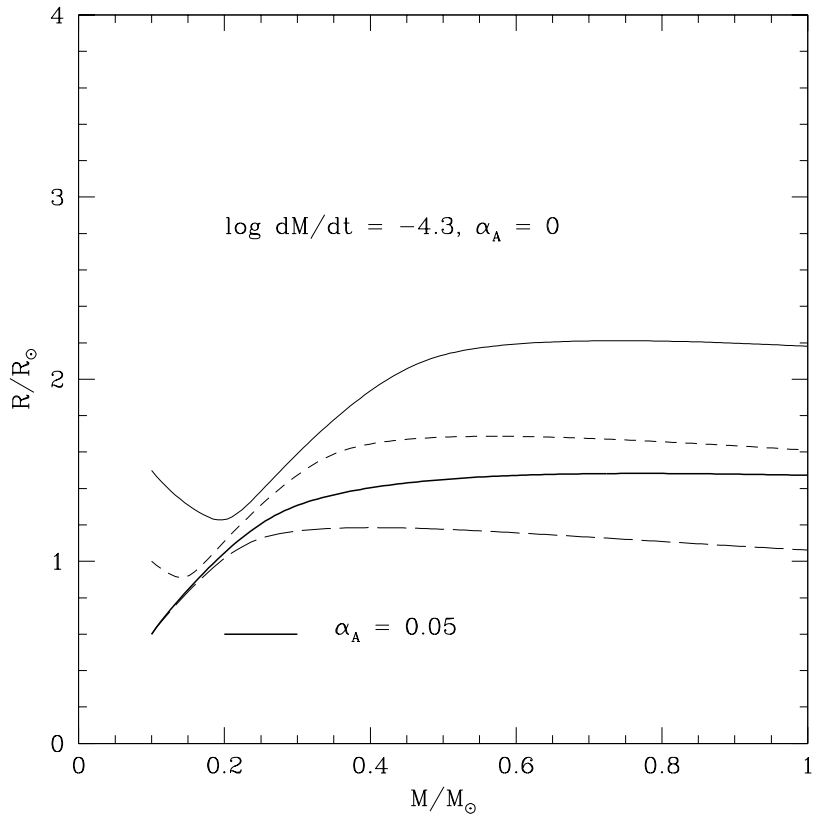}
\caption{
Simple polytropic model for rapid cold disk accretion, varying 
the initial radius, along with one ``warm accretion'' model
(heavy curve) (see text)}
\label{fig:fig3}
\end{figure}

Figure \ref{fig:fig3} shows protostellar $R(M)$ calculated as above for a variety of
initial radii for a rapid mass accretion rate of $\mdot = 5 \times 10^{-5} \msunyr$
as in BCG and BC.  We begin with an initial core mass of $0.1 \msun$ because
our polytropic approximation is not adequate at lower masses.
It is evident that the initial radius is an extremely important parameter
in establishing the subsequent evolutionary track. 
Starting with a relatively large initial radius
allows the protostar to evolve for a period of time along the D fusion
``birthline'' (where the central temperature is $\sim 1 \times 10^6$~K,
the point at which D fusion can begin),
while for smaller starting radii the protostar never moves
up to the birthline beyond $0.2 \msun$, but stays well below.

The reason for this behavior is straightforward.  For the star not to
contract under cold accretion, 
the energy available from the fusion of D must be sufficient to
compensate for the increased gravitational potential energy due to the
accreted mass.  At a given mass, smaller radii imply larger gravitational 
potential energies, which D fusion energy release cannot counteract in steady 
state.  Our simple model assumes an initial full complement
of deuterium, instantaneous mixing of D throughout
the star, and that the star remains an $n = 1.5$ polytrope.
With these assumptions, the star initially expands, moving up along the
birthline, by fusing D faster than it is being accreted.  This behavior is forced
by the high temperature sensitivity of D fusion (Stahler 1988), which prevents
the central temperature from increasing very much beyond the point at
which D begins to be fused.  The initial phase of
expansion rapidly exhausts the initial D content, using the released energy
to expand the star such that $T_c \propto G M/R \sim$~constant.  
After the D is exhausted, the
accretion of fresh material cannot keep up with the photospheric radiative losses,
and the protostar adds mass at nearly constant or slightly decreasing radius.

The BCG and BC models do not show the initial rise from $0.6 \rsun$ to 
$\sim 1 \rsun$ seen in our calculation because the accretion turns on and
off on timescales much shorter than the Kelvin timescale. 
However, the sense of the evolutionary behavior is the same,
in the sense of producing much smaller protostellar radii.  While BCG and BC
emphasize episodic accretion, they point out that this is not essential;
steady fast accretion yields the same basic results (see \S 5).  

Figure \ref{fig:fig3} also shows results for $\alpha_A = 0.05$, that is, 5\% of the accretion
luminosity is assumed to be added to the star as thermal energy, for the case of
the smallest initial radius ($0.6 \rsun$) (heavy curve).  The results show a significant
increase in the final protostellar radius.  Such values of added thermal energy
are not implausible; for instance, we estimate that for typical mass to radius
protostellar values, inner disk temperatures $\sim 1 \times 10^5$~K yield
$\alpha_A \approx 0.03$.  However, it must be re-emphasized that the use of
equation (\ref{eq:laccd}) at high accretion rates
in any case is suspect, as it does not include any
irradiation of the star by the disk.  As noted above, the net effect of a large
external accretion luminosity is likely to turn the outer layers of the star radiative
and result in expansion (as occurs in Prialnik \& Livio 1985 for sufficient heat addition
during accretion).

\section{Discussion}

The post-protostellar HR diagram positions of the rapid cold accretion models 
of BCG and BC lie well below observations of young stars in many molecular clouds,
as BCG themselves show (their Figure 1; see also, e.g., Rebull \etal 2002).
They also lie below estimated protostellar positions 
(White \& Hillenbrand 2004; Doppmann \etal 2005; see Figure 4 in White \etal 2007), though 
one must emphasize that these results are somewhat uncertain due to the difficulty
of making appropriate extinction corrections.
Moreover, some young clusters of age $\sim 5-10$~Myr show quite narrow
empirical isochrones (Moitinho \etal 2001; Preibisch \etal 2002
Hernandez \etal 2007, 2008);
this would not be the case if a substantial fraction of the stars
exhibited offsets of several Myr in their isochrones.  In all,
the observational constraints on young stellar populations indicate
that age uncertainties due to rapid accretion are generally
considerably less than the maximum values suggested by BCG and BC.

On the other hand, this is not to understate the importance of the BCG and BC
results in emphasizing the importance of initial protostellar radii
( ``protostellar core radii'') for subsequent protostellar evolution.  
These radii are controlled by the amount of thermal energy incorporated
into the initial protostellar core during the early phases of
hydrodynamic collapse, and which, as discussed in the Introduction, are difficult to calculate.
Given the likely variation in initial conditions for
collapse (as indicated by models of turbulent star formation), there is no
reason to suppose that all stars in a given region started out with the same
initial core mass and radius, or even all stars of the same mass.  

\begin{figure}
\includegraphics[angle=0,width=0.47\textwidth]{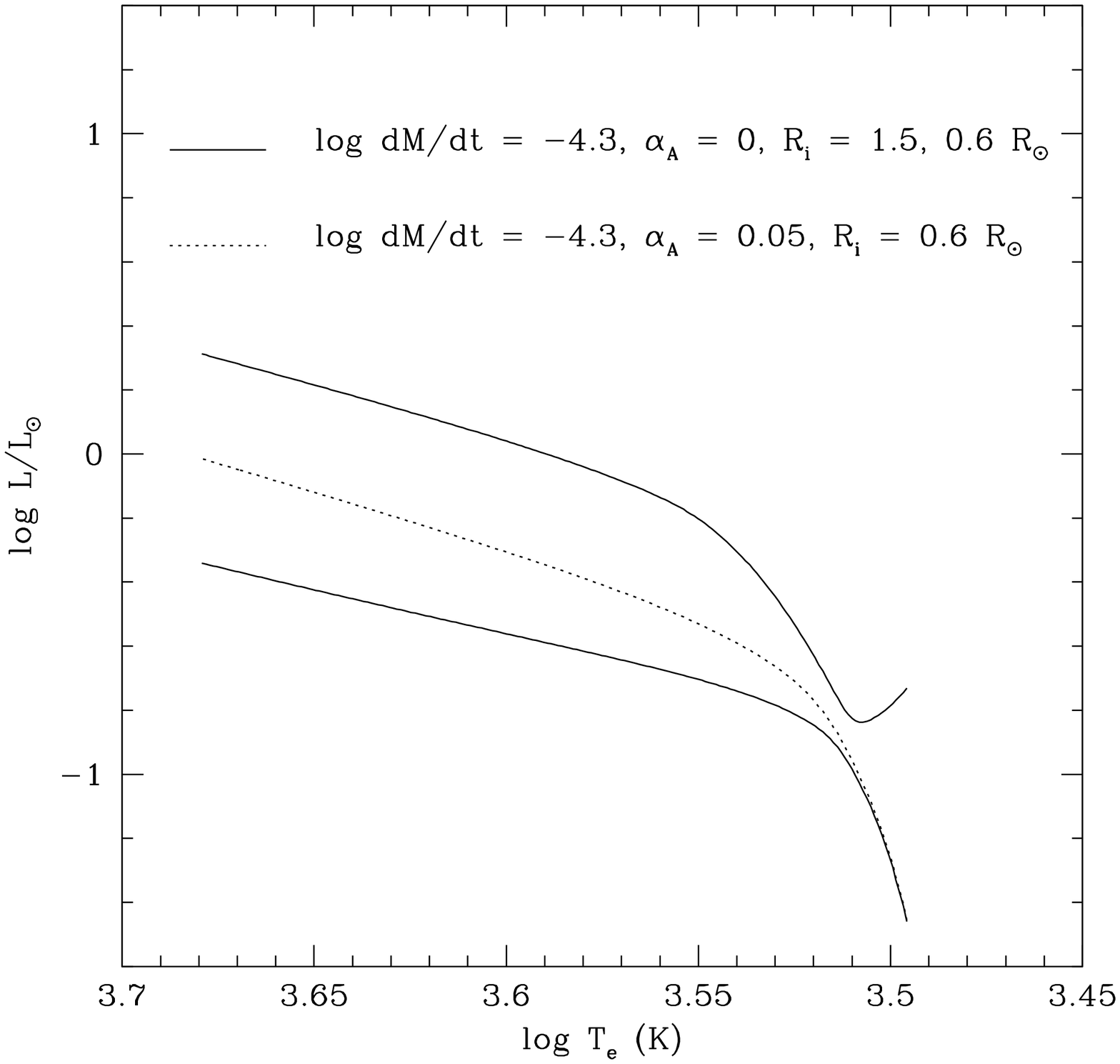}
\includegraphics[angle=0,width=0.47\textwidth]{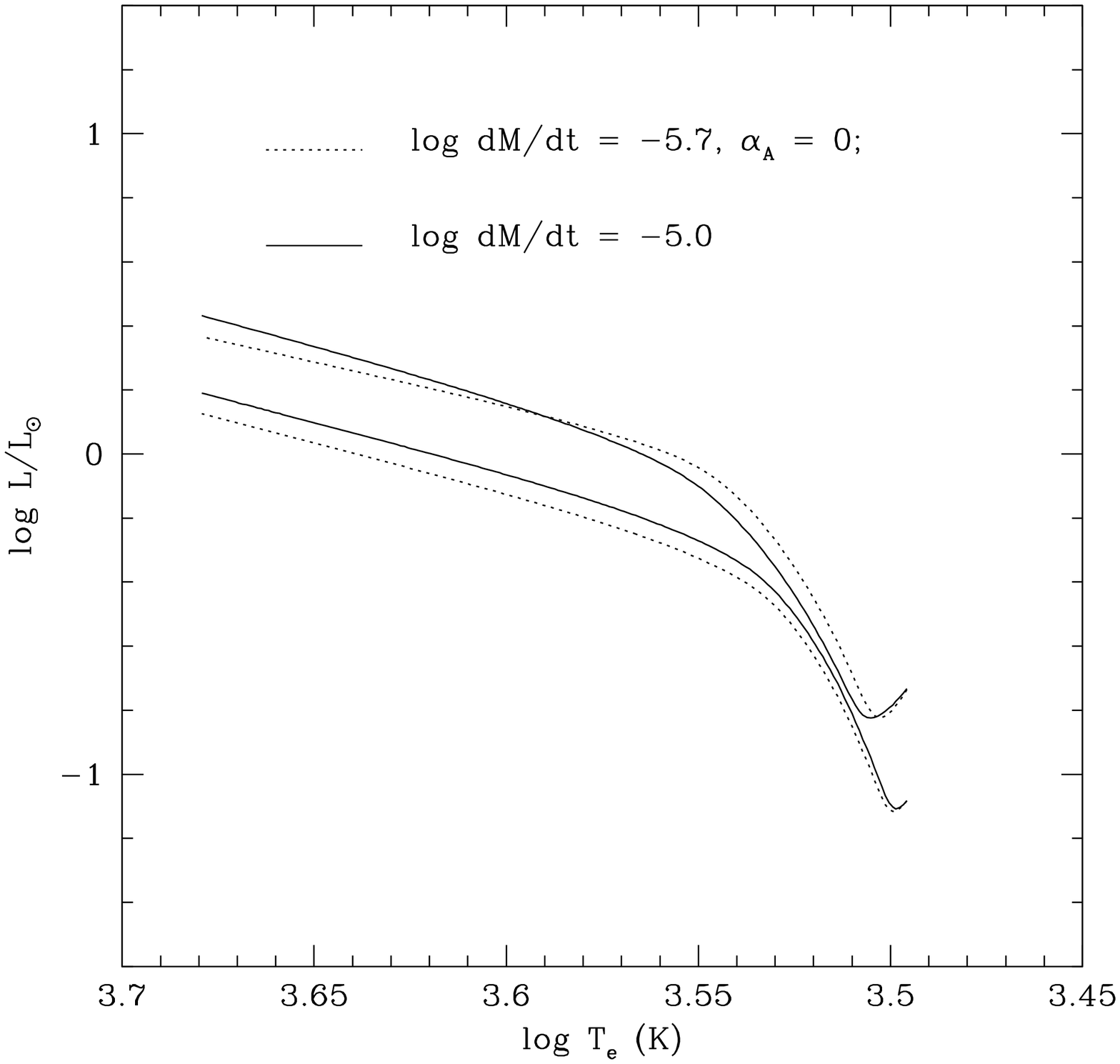}
\caption{
Left panel: Evolutionary tracks for the models shown in Figure \ref{fig:fig3}
(using the same symbols to differentiate the cases).
Right panel: Evolutionary tracks for slower cold accreting models.
The results show the importance of the initial assumed radius, even
in more traditional, slower-accretion evolution (see text)}
\label{fig:fig4}
\end{figure}

That variations in initial radii would result in protostars evolving
in ``birth regions'' rather than strict birthlines has long been recognized,
even in the case of steady accretion (see Figure 6 in HCK).  
To emphasize this point, in the right-hand
panel of Figure \ref{fig:fig4} we show steady cold accretion tracks for the
slower rates more appropriate to the use of equation \ref{eq:laccd}.
The results indicate that the initial ``core'' radius dominates the results, with hardly
any difference between accretion rates of $2 \times 10^{-6}$ and $10^{-5} \msunyr$.
To place this in perspective, contraction from large
radii of a fully convective star down vertical tracks in the HR diagram 
is such that the age $t \propto L_*^{-3/2}$ (e.g., Hartmann 2009).  The 
difference of about $0.2$ in $\log L_*$ shown in Figure \ref{fig:fig4}
therefore corresponds to a difference in $\log t \sim 0.3$, or a change in the apparent
age of a factor of two.  
(Of course, the age of the star immediately after the end of accretion is effectively zero).
Although there are systematic uncertainties in comparing the results of this
simple calculation to those of more sophisticated treatments, it is
still worth noting that the two sets of evolutionary tracks for protostars with
$\log T_{eff} = 3.6$ correspond to ages of $\sim 1$ and $2$~Myr in
the Siess \etal (2000) tracks, as the latter have no birthline corrections.  
This suggests that uncertainties in ages could be of order 1 Myr for the low-mass
stars; because D fusion is less important in intermediate mass stars, age
uncertainties can be larger for these objects (Hartmann 2003).
The need for birthline or birth region
corrections to ages computed from non-accreting tracks, and its importance at
early times, was the reason why Baraffe \etal (1998) did not provide
evolutionary tracks and isochrones for ages less than 1 Myr (personal communication).   

It is understandable that observers often do not include birthline corrections in
determining age distributions, and instead use tracks assuming contraction from
much larger radii, to avoid a number of problems.  
Birthline positions are theoretically uncertain, 
and might even be variable from star to star.  In addition, 
there are significant uncertainties due to systematic effects in translating 
theoretical tracks into observable quantities.  However, this does not mean 
that birth region corrections are unimportant.
As BCG and BC suggest, this could have important implications for some of the
variations in radius seen in among stars of the same mass in
young clusters (e.g., Jeffries 2007, 2011) and in eclipsing young brown dwarf binaries
(Stassun \etal 2006; Mohanty \etal 2009).

A robust calculation of the amount of thermal energy carried into a protostar
by a rapidly-accreting thick disk is an enormously challenging problem,
requiring three-dimensional simulation and careful treatment of viscosities in
both disk and star.  A more immediately feasible theoretical calculation,
following the methods of BCG and BC, would be to include a plausible
amount of thermal energy addition in rapid accretion burst models to see 
this can explain the factor of $\sim$ two increase in stellar radius implied 
by the spectral energy distribution fitting of FU Ori objects.
Unfortunately, neither of these theoretical improvements will address the
complex issue of what variations in initial protostellar core radii arise
from hydrodynamic collapse in turbulent, structured molecular clouds.
As it seems completely unreasonable to suppose that protostellar cores will all have the
same starting masses and radii - or even the same starting radii, given a fixed
mass - this variation will result in differing pre-main sequence radii at a given
mass and true age after the end of major accretion.

On theoretical and observational grounds, rapid cold disk
accretion is highly unlikely to occur; thus we argue that rapid episodic 
accretion is not a good candidate for producing very large 
variations in protostellar radii, with consequent uncertainties of $\sim$~10 Myr  
in the ages of low-mass pre-main sequence stars.  More modest uncertainties,
of order 1 Myr (and more for intermediate mass stars; Hartmann 2003) are
plausible, and can be produced by differences in initial conditions, most particularly
the initial protostellar core radii; this can probably only be advanced
through improved observational constraints.
The results of BCG and BC emphasize the importance of further efforts to
improve our understanding of the ages of the youngest stars.

\acknowledgments

We acknowledge useful conversations with Keith MacGregor, Isabelle Baraffe,
\& Gilles Chabrier, and a helpful report from an anonymous referee.
This work was supported in part by NASA grant NNX08A139G and by the
University of Michigan. 

\section{Appendix}

At high accretion rates, viscous heating dominates the disk temperature distribution.
Vertically-averaging the disk, the relationship between the central temperature
$T_c$ and the effective temperature $T_{eff}$ is
$T_c^4 = 3/16 T_{eff}^4 \Sigma \kappa$, where 
$\Sigma$ is the disk surface density and $\kappa$ is the Rosseland mean opacity.
Fitting the opacity as $\kappa = \kappa_{r} (T_c/T_r)^{\alpha}\, 
(P_c/P_r)^{\beta}$, where $T_r$ and $P_r$ are constants,
and combining this with the 
steady disk results $T_{eff}^{4} = 3 G M \dot{M} /8 \pi R^3 \sigma$ and
$\alpha c_{s} H \Sigma = \dot{M} /3 \pi$ to find
the disk central pressure,
\begin{equation}
P_c= {k \over \mu m_{u}} \rho T_{c}
=\frac{\dot{M}}{6\pi\alpha c_{s}H^{2}}\frac{k T_{c}}{\mu m_{u}}\,. 
\end{equation}
The disk central temperature is
\begin{equation}
(\frac{R}{R_{0}})^{9/2}=\frac{3}{16}(\frac{T_{r}}{T_{c}})^{5-\alpha}
\frac{3GM\dot{M}}{8\pi \sigma R_{0}^{3} T_{r}^{4}}
\frac{\dot{M}}{3\pi\alpha}\frac{\mu m_{u}\kappa(GM)^{1/2}}{k R_{0}^{3/2} T_{r}}\,.
\end{equation}
Combing above two equations, we have
\begin{eqnarray}
\left ({R \over R_0} \right)^{9/2+3\beta} & = & 
{3 \over 16} \left ( {T_r \over T_c} \right)^{5-\alpha+\beta/2} {3 G M \dot{M} \over 8 \pi \sigma R_0^3 T_r^4}\nonumber \\
& & {\dot{M} \over 3\pi\alpha}
{\mu m_u\kappa_0(GM)^{1/2} \over k R_0^{3/2} T_{r}}\nonumber \\
& & \left ({\dot{M}GM \over R_0^{3} 6 \pi \alpha(kT_r/\mu m_u)^{1/2} P_r} \right)^{\beta}\label{eq:Tc}
\end{eqnarray}
The disk becomes geometrically thick when the thermal instability develops,
which occurs for central temperatures above 5000~K.  Setting
$\beta =0.4$, $\kappa_0 =0.1$ at $T_{c}=T_{r}=5000 K$, and $\log P_r \sim 3$, we have
\begin{equation}
R_{5000} = 0.46 \left ( {M \over M_{\odot}} \right)^{0.33}
\left ({\dot{M} \over 10^{-4}\msunyr} \right )^{0.42}
\left ({\alpha \over 0.1} \right)^{-0.25} {\rm AU} \,.
\end{equation}
Comparison with the two-dimensional simulations of Zhu \etal (2009b) indicates
that the disk becomes geometrically thick at about a factor of two smaller
radius, where the central temperature rises above $\sim 2 \times 10^4$~K.
Adopting this scaling, we arrive at an estimate of the radius interior to
which the disk is geometrically thick and internally hot as
\begin{equation}
R_{T} \sim 0.23 \left ( {M \over M_{\odot}} \right)^{0.33}
\left ({\dot{M} \over 10^{-4}\msunyr} \right )^{0.42}
\left ({\alpha \over 0.1} \right)^{-0.25} {\rm AU} \,.
\end{equation}

\end{document}